\documentclass{aa}
\usepackage{graphics}

\begin{document}

\thesaurus{ 3(
        11.01.2; 
        11.19.3; 
        13.09.1) 
        }

\title
{Mid-infrared diagnostics to distinguish AGNs from starbursts
\thanks{Based on observations made with ISO, an ESA project with
instruments funded by ESA Member States (especially the PI countries:
France, Germany, the Netherlands and the United Kingdom) and with the
participation of ISAS and NASA.}}

\author{O. Laurent\inst{1,2},
        I.F. Mirabel\inst{1,3},
        V. Charmandaris\inst{4,5}, 
        P. Gallais \inst{1},
        S.C. Madden \inst{1},
        M. Sauvage\inst{1}, 
        L. Vigroux\inst{1}
        \and 
        C. Cesarsky \inst{1,6}}

\institute{
CEA/DSM/DAPNIA Service d'Astrophysique F-91191 Gif-sur-Yvette, France
\and
Max-Planck-Institut f\"ur extraterrestrische Physik, Postfach 1603, 85740 Garching, Germany
\and
Instituto de Astronom\'\i a y F\'\i sica del Espacio. cc 67, 
suc 28. 1428 Buenos Aires, Argentina
\and 
Observatoire de Paris, DEMIRM, 61 Av. de l'Observatoire, 
F-75014 Paris, France
\and
Astronomy Department, Cornell University, Ithaca NY, 14853, USA
\and
European Southern Observatory, 85748 Garching, Germany
}

\titlerunning{Mid-infrared diagnostics to distinguish AGNs from 
starbursts}

\authorrunning{Laurent et al.}

\date{Received 1 June 1999 / Accepted 18 May 2000}

\offprints{O. Laurent, olaurent@mpe.mpg.de} 

\maketitle

\begin{abstract}
We present new mid-infrared (MIR) diagnostics to distinguish emission
of active galactic nuclei (AGN) from that originating in starburst
regions. Our method uses empirical spectroscopic criteria based on the
fact that MIR emission from star forming or active galaxies arises
mostly from HII regions, photo-dissociation regions (PDRs) and AGNs.  The
analysis of the strength of the 6.2\,$\mu$m Unidentified Infrared Band
(UIB) and the MIR continuum shows that UIBs are very faint or absent
in regions harboring the intense and hard radiation fields of AGNs and
pure HII regions, where the UIB carriers could be destroyed. The
MIR signature of AGNs is the presence of an important continuum in the
3-10\,$\mu$m band which originates from very hot dust heated by the
intense AGN radiation field. Using these two distinct spectral
properties found in our MIR templates, we build diagnostic diagrams which
provide quantitative estimates of the AGN, PDR and HII region
contribution in a given MIR spectrum. This new MIR classification can
be used to reveal the presence of AGNs highly obscured by large columns
of dust.

\keywords{ Galaxies: active --
        Galaxies: starburst --
        Infrared: galaxies}

\end{abstract}

\section{Introduction}

Since the discovery by IRAS of ultraluminous infrared galaxies
(L(8-1000\,$\mu$m) $>$ 10$^{12}$L$_\odot$) which emit the bulk of
their energy in the infrared, a large number of studies have shown that
intense star forming regions as well as AGNs are necessary to explain
these high luminosities (see \cite{Sanders2} for a review). Large
concentrations of molecular gas are needed for fueling nuclear
starbursts and/or AGNs. Consequently, absorption makes the
distinction between starburst and AGN activity difficult and the
estimate of their relative contribution to the total infrared
luminosity is far from straightforward. Optical classification
(\cite{Veilleux}, \cite{Goldader}, \cite{Kim}) is plagued by
extinction, while radio wavelengths, although free from extinction, do
not result in a definite classification (\cite{Condon}, \cite{Lonsdale},
\cite{Smith}). Based on X-ray observations (\cite{Ogasaka}, \cite{Risaliti}), 
ground-based infrared observations (\cite{Roche}, \cite{Dudley2},
\cite{Murphy}, \cite{Soifer}) and on ISO observations
(\cite{Genzel}, \cite{Lutz3}, \cite{Rigopoulou}), considerable
progress has been made in defining the fraction of the AGN/Starburst
contributions to the bolometric luminosity of ultraluminous infrared
galaxies.

To further examine the AGN and starburst connection, we have observed
a sample of nearby active and interacting galaxies harboring starburst
regions and/or AGNs using ISOCAM. Due to the unique spectro-imaging
capabilities of the camera which provide an angular resolution of
4-8\,arcsec between 5 and 16\,$\mu$m (200-400\,pc at 10\,Mpc), we can
reveal and study obscured central regions not visible at optical
wavelengths (A$_{15\,\mu m}$$\sim$\,A$_{V}$/70, \cite{Mathis}).

The goal of this paper is to present an empirical method to
distinguish and quantify the MIR emission coming from starbursts
and AGNs. Our approach is based on a new MIR diagnostic for estimating
the relative importance of these two main energy sources found in
galactic centers.  Our sample and the data reduction methods are
described in section 2. A brief summary of the typical MIR emission
encountered in these galaxies is presented in section 3, while our MIR
templates and diagnostic diagrams are shown in section 4. We discuss
our results and their implications in section 5 and the final
conclusions are given in section 6.

\section{Observations and data reduction}

A large sample of nearby galaxies hosting star formation activity and
AGN signatures (see Table \ref{sample}) was observed with the ISOCAM
camera (\cite{Cesarskya}) on board the Infrared Space Observatory
(ISO, \cite{Kessler}). All observations come from the ISOCAM
consortium guaranteed time programs and most of them were part of the
active galaxy proposal CAMACTIV (P.I. I.F. Mirabel,
e.g. \cite{Charmandaris1}, \cite{Mirabela}, \cite{Gallais},
\cite{Mirabelb}). These galaxies have been observed either in the
spectro-imaging mode with the Circular Variable Filter (CVF) or in the
raster mode with broad band filters.

The CVF covers a spectral range from 5 to 16\,$\mu$m with a spectral
resolution of $\sim$\,40.  For CVF scans, a single pointing was made
for each galaxy, using 1.5-3\,arcsec/pixels, giving a total field of
view for the 32$\times$32 pixel array of 0.8-1.6\,arcmin. The spatial
resolution is 4-8\,arcsec limited by the full width at half maximum
(FWHM) of the point spread function (PSF). High spatial resolution and
sensitivity are essential in our analysis in order to isolate the
central regions in nearby galaxies. Approximately 12 exposures of 2-5s
each were made at each wavelength in addition to $\sim$\,20
exposures at the start of each observation to decrease the initial
detector transient effects.

The raster maps were made with various broad band filters designed to
select distinct features such as the continuum emission of the very
small grains using the LW3(12-18\,$\mu$m) filter, and the UIBs with
the LW2(5-8.5\,$\mu$m) and LW4(5.5-6.5\,$\mu$m) filters. Depending on
the apparent size of the galaxy, a pixel field of view of 1.5 or 3\,arcsec was
selected. Most of the maps were 2$\times$2 rasters with 6 pixel
overlap consisting of approximately 200 exposures of 2-5s
each. Another 30 exposures were added at the start of the observations
to decrease the effects of transients.

The ISOCAM data were analyzed with the CAM Interactive Analysis
software (CIA\footnote{CIA is a joint development by the ESA
astrophysics division and the ISOCAM consortium led by the ISOCAM PI,
C. Cesarsky, Direction des Sciences de la mati\`ere, C.E.A. France.})
and were calibrated with the general methods described in
\cite{Starck3}. To correct for the dark current, a dark model taking
into account the observing time parameters was subtracted. Cosmic ray
hits were removed by applying a multiresolution median filtering
(\cite{Starck1}).  Corrections of detector memory effects were made
with the so-called inversion method (\cite{Abergel}). The flat field
correction was performed using the library of calibration
data. Finally, individual exposures were combined using shift
techniques in order to correct the jitter effect due to the satellite
motions (amplitude $\sim$ 0.5\,arcsec).  To derive the photometry of
individual galaxy regions, aperture corrections as well as
deconvolution techniques (\cite{Starck2}), were applied to account for
the overall extension of the PSF in a few cases. We estimate that the
absolute uncertainty of our photometric measurements is $\sim$\,20$\%$
while the error in the relative uncertainty mainly due to errors on
transient effect correction is $\sim$10$\%$.

\section{MIR emission of normal, starburst galaxies and AGNs.}

MIR spectra (5-16\,$\mu$m) in galaxies can mainly arise from a variety
of physical components including: 1) the evolved stellar population
(Rayleigh-Jeans regime), which can dominate in early-type galaxies, 2)
emission from the ionized interstellar gas, 3) non-thermal emission
from radio sources and 4) dust particules responsible for the
underlying continuum and 5) carriers of UIBs centered at 6.2, 7.7,
8.6, 11.3 and 12.7\,$\mu$m.  The UIBs are thought to be due to C=C and
C--H stretching and bending vibration modes in carbonaceous materials
(\cite{Leger1}, \cite{Allamandola}, \cite{Papoular}) and they
dominate the MIR spectra of galaxies with low or intermediate star
formation activity (\cite{Mattila2}) as in PDRs (\cite{Klein}) and
diffuse galactic regions (\cite{Giard}, \cite{Mattila1}).  The dust
emitting longward of about 10\,$\mu$m, is attributed to Very Small
Grains (here-after VSGs) with radius less than 10\,nm
(\cite{Desert1}), and is prominent in regions actively forming stars,
such as Galactic HII regions (e.g. \cite{Verstraete},
\cite{Cesarskyc}) and in starburst regions. To emit in the MIR, 
the carriers of the UIBs and the VSGs are thought to be stochastically
heated by the stellar radiation field reaching temperature
fluctuations of the order of 100-1000 K (\cite{Puget},
\cite{Desert1}). The emission from ionized gas is mainly observed
through forbidden lines such as [ArII](6.9\,$\mu$m),
[NeVI](7.6\,$\mu$m), [ArIII](8.9\,$\mu$m), [SIV](10.5\,$\mu$m),
[NeII](12.8\,$\mu$m), [NeV](14.3\,$\mu$m) and
[NeIII](15.6\,$\mu$m). Because of their high ionization potential,
these lines can be used to trace the hardness of the radiation field.
The presence of [NeVI] and [NeV] has also provided evidence for AGNs
(\cite{Moorwood}, \cite{Genzel}, \cite{Sturm}) or supernova remnants 
(\cite{Oliva}).

In order to understand how the total MIR emission in our sample of
starbursts and AGNs varies according to these different contributions,
we first review the well-studied MIR properties in resolved nearby 
galaxies from early-type to later type disk galaxies and then address
our observations in light of these properties.

\newpage
\begin{table*}[!t]
\caption[]{Our galaxy sample}
\begin{tabular}{lrrccccccc}   
\hline
Source                  & RA (J2000)              &    DEC (J2000)                           & LW2     & LW3    & LW4        & \underline{LW3}   & \underline{LW2}   & \multicolumn{2}{c}{Spectral type}\\
                   &                    &                                  & (mJy)  & (mJy)      & (mJy) & LW2          & LW4      
& LW & CVF\\
\hline
NGC\,1068(Nuc\,:\,9\,'')(2)           & 02$^h$ 42$^m$ 40.6$^s$    & -00$\degr$ 00$\arcmin$ 47.8$\arcsec$             & 13009  & 46616       & 10221 & 3.58    & 1.27 & AGN & AGN\\
Arp\,118(Nuc\,:\,3\,'')(4)       & 02$^h$ 55$^m$ 12.2$^s$    & -00$\degr$ 11$\arcmin$ 00.8$\arcsec$       & 16      & 42          & 10         & 2.62    & 1.62 & AGN & AGN\\
NGC\,3147(Nuc\,:\,9\,'')(5)$^{\dag}$   & 10$^h$ 16$^m$ 53.6$^s$    &  73$\degr$ 24$\arcmin$ 03.3$\arcsec$                            & 8     & 22          & 9          & 2.60    & 0.90 & \hspace{2.1mm}AGN$^{+}$ & --\\
Centaurus A(Nuc\,:\,4.5\,'')(6)   & 13$^h$ 25$^m$ 27.6$^s$     & -43$\degr$ 01$\arcmin$ 08.8$\arcsec$                                & 575        & 1658   & 500        & 2.89    & 1.15 & AGN & AGN\\
NGC\,6814(Nuc\,:\,9\,'')(7)$^{\dag}$  & 19$^h$ 42$^m$ 40.6$^s$    & -10$\degr$ 19$\arcmin$ 24.6$\arcsec$                               & 33       & 85          & 31    & 2.59    & 1.05 & AGN & --\\
\hline
\\
\hline
NGC\,253(Nuc\,:\,7.5\,'')(6)          & 00$^h$ 47$^m$ 33.1$^s$    & -25$\degr$ 17$\arcmin$ 17.8$\arcsec$         & 4703   & 15716      & 2296  & 3.34    & 2.05  & HII & HII\\
Arp\,236(Nuc\,:\,4.5\,'')(3)          & 01$^h$ 07$^m$ 47.5$^s$    & -17$\degr$ 30$\arcmin$ 25.6$\arcsec$       & 200     & 358    & 108        & 1.79    & 1.84 & PDR & \hspace{1.5mm}AGN$^{*}$\\
NGC\,1808(Nuc\,:\,9\,'')(1)           & 05$^h$ 07$^m$ 42.3$^s$    & -37$\degr$ 30$\arcmin$  46.1$\arcsec$                       & 1074   & 1450       & 538        & 1.35    & 2.00 & PDR & PDR\\
M82(Nuc\,:\,9\,'')(6)      & 09$^h$ 55$^m$ 52.2$^s$    &  69$\degr$ 40$\arcmin$ 46.9$\arcsec$        & 5198   & 12720      & 2573  & 2.45    & 2.02 & HII & HII\\
NGC\,3256(Nuc\,:\,4.5\,'')(1)         & 10$^h$ 27$^m$ 51.8$^s$    & -43$\degr$ 54$\arcmin$ 08.7$\arcsec$                             & 442        & 1196   & 212        & 2.70    & 2.09 & HII & HII\\
Arp\,299(A\,:\,4.5'')(1)             & 11$^h$ 28$^m$ 34.2$^s$    &  58$\degr$ 33$\arcmin$ 46.5$\arcsec$       & 325      & 1860   & 108        & 5.73    & 3.00 &\hspace{2.1mm}HII$^{+}$ & \hspace{2.1mm}HII$^{+}$\\
Arp\,299(B\,:\,4.5\,'')(1)           & 11$^h$ 28$^m$ 31.5$^s$    &  58$\degr$ 33$\arcmin$ 40.4$\arcsec$       & 505      & 1951   & 303        & 3.86    & 1.67 & HII & HII\\
Arp\,299(C'\,:\,4.\,5'')(1)          & 11$^h$ 28$^m$ 31.8$^s$    &  58$\degr$ 33$\arcmin$ 49.9$\arcsec$       & 76       & 232    & 36         & 3.06    & 2.14 & HII & HII\\
Arp\,299(C\,:\,4.5\,'')(1)           & 11$^h$ 28$^m$ 31.2$^s$    &  58$\degr$ 33$\arcmin$ 48.9$\arcsec$       & 126      & 461    & 65         & 3.66    & 1.93 & HII & HII\\
NGC\,4038(KnotA\,:\,6\,'')(8)    & 12$^h$ 01$^m$ 54.9$^s$    & -18$\degr$ 53$\arcmin$ 12.3$\arcsec$       & 23      & 135    & 13    & 5.78    & 1.75 & HII & HII\\
Arp\,220(9)           & 15$^h$ 34$^m$ 57.2$^s$    &  23$\degr$ 30$\arcmin$ 10.8$\arcsec$       & 162      & 732    & 79         & 4.50    & 2.07 & HII & HII\\
NGC\,6240(9)        & 16$^h$ 52$^m$ 58.5$^s$    &  02$\degr$ 24$\arcmin$ 03.4$\arcsec$       & 229      & 758    & 107        & 3.30    & 2.14 & HII & HII\\
IRAS\,17208-0014(1)$^{\dag}$         & 17$^h$ 23$^m$ 21.9$^s$    & -00$\degr$ 17$\arcmin$ 00.4$\arcsec$       & 127      & 248    & 60         & 1.96    & 2.11 & PDR & --\\
IRAS\,19254-7245(1)$^{\dag}$        & 19$^h$ 31$^m$ 21.6$^s$    & -72$\degr$ 39$\arcmin$ 20.2$\arcsec$       & 111       & 264    & 84    & 2.37    & 1.33 & \hspace{1.5mm}AGN$^{*}$ & --\\
IRAS\,20551-4250(1)$^{\dag}$  & 20$^h$ 58$^m$ 26.8$^s$    & -42$\degr$ 39$\arcmin$ 00.6$\arcsec$       & 123        & 425    & 62    & 3.47    & 1.99 & HII & --\\
IRAS\,23128-5919(1)   &  23$^h$ 15$^m$ 47.0$^s$      & -59$\degr$ 03$\arcmin$ 14.0$\arcsec$       & 90    & 319    & 53    & 3.54    & 1.70 & HII & HII\\
\hline
\\
\hline
NGC\,253(Disk)(1)      & 00$^h$ 47$^m$ 33.1$^s$    & -25$\degr$ 17$\arcmin$ 17.8$\arcsec$        & 339     & 681    & 171         & 2.00    & 1.98 & PDR & PDR\\
NGC\,520(1)           & 01$^h$ 24$^m$ 34.8$^s$    &  03$\degr$ 47$\arcmin$ 30.8$\arcsec$       & 486      & 511    & 231        & 1.05    & 2.10 & PDR & PDR\\
NGC\,1068(Disk)(2)              & 02$^h$ 42$^m$ 40.6$^s$    & -00$\degr$ 00$\arcmin$ 47.8$\arcsec$                          & 246       & 296    & 137       & 1.20     & 1.80 & PDR & PDR\\
NGC\,1808(Disk)(1)        & 05$^h$ 07$^m$ 42.3$^s$    & -37$\degr$ 30$\arcmin$  46.1$\arcsec$                    & 1615   & 2694      & 728        & 1.67    & 2.22 &  \hspace{2.1mm}PDR$^{+}$ & PDR\\
M\,82(Disk)(1)     &09$^h$ 55$^m$ 52.2$^s$    &  69$\degr$ 40$\arcmin$ 46.9$\arcsec$          & 1177   & 1133      & 589        & 0.96    & 2.00 & PDR & PDR\\
NGC\,3147(5)$^{\dag}$       &10$^h$ 16$^m$ 53.6$^s$    &  73$\degr$ 24$\arcmin$ 03.3$\arcsec$     & 375        & 483    & 215        & 1.29    & 1.74 & PDR & --\\
NGC\,3256(Disk)(1)             &10$^h$ 27$^m$ 51.8$^s$    & -43$\degr$ 54$\arcmin$ 08.7$\arcsec$          & 202       & 420    & 85         & 2.07    & 2.38 & \hspace{2.1mm}PDR$^{+}$ & \hspace{2.1mm}PDR$^{+}$\\
NGC\,3263(Nuc\,:\,4.5\,'')(1)$^{\dag}$      & 10$^h$ 29$^m$ 13.1$^s$    & -44$\degr$ 07$\arcmin$ 22.0$\arcsec$        & 59    & 62          & 31         & 1.03    & 1.94 & PDR & --\\
NGC\,4676(A)(1)$^{\dag}$    & 12$^h$ 46$^m$ 10.1$^s$     &  30$\degr$ 43$\arcmin$ 57.2$\arcsec$       & 58     & 30          & 40         & 0.53    & 1.44 & \hspace{2.1mm}PDR$^{+}$ & -- \\
NGC\,4676(B)(1)$^{\dag}$    & 12$^h$ 46$^m$ 11.4$^s$     &  30$\degr$ 43$\arcmin$ 23.1$\arcsec$      & 4.11       & 2.15           & 2.17          & 0.52       & 1.89 & PDR & --\\
Centaurus A(Disk)(10)  &13$^h$ 25$^m$ 27.6$^s$     & -43$\degr$ 01$\arcmin$ 08.8$\arcsec$               & 62    & 101    & 35         & 1.62    & 1.78 & PDR & PDR\\
NGC\,6814(6)$^{\dag}$       &19$^h$ 42$^m$ 40.6$^s$    & -10$\degr$ 19$\arcmin$ 24.6$\arcsec$         & 291      & 290    & 144   & 1.00    & 2.02 & PDR & --\\
NGC\,7252(1)            & 22$^h$ 20$^m$ 44.9$^s$    & -24$\degr$ 40$\arcmin$ 41.3$\arcsec$       & 142     & 185    & 67    & 1.30    & 2.11 & PDR & --\\
\hline\\
\end{tabular}
\label{sample}
{\bf Table note:} Our galaxy sample is presented in three parts.  The
top includes nuclei of galaxies containing an AGN, the middle contains
regions harboring starburst activity and the bottom presents quiescent
star forming regions.  The spectral type column indicates which of the
three components (HII, PDR or AGN) provides the dominant
contribution. Two values, one according to our broad band diagnostic
(left column, see also Fig. \ref{fig8}) and one based on the CVF
diagnostic (right column, see also Fig.\ref{fig6}) are presented.  The
broad band fluxes for most galaxies have been calculated from their
CVF spectra.  NGC\,7252 was observed in the CVF mode but due to its
weak continuum emission the errors of the CVF diagnostic are large so
we include only the LW classification. In NGC\,4676(A), the MIR
spectral energy distribution is probably contaminated by stellar
emission (\cite{Hibbard}, \cite{Read}).\\ Notation used: ``Nuc''=
nucleus with the diameter in arcsec, ``Disk''= well detected star
formation region in the disk. References: $^{1}$ This work; $^{2}$
\cite{Laurent2}; $^{3}$ \cite{Laurent4}; $^{4}$ \cite{Charmandaris2}; 
$^{5}$ \cite{Laurent1}; $^{6}$ \cite{Laurent5}; $^{7}$
\cite{Laurent3}; $^{8}$ \cite{Mirabela}; $^{9}$ \cite{Charmandaris1}; $^{10}$
\cite{Mirabelb}. The identification of specific regions in Arp\,299,
NGC\,4038 and NGC\,4676 follows the usual notation found in the
literature.  Galaxies marked with\,$^{\dag}$ have been observed only
in broad band filter mode (LW). Galaxies marked with $^{+}$ identify
those which clearly fall outside the limits of our diagnostic, as
shown in Figure \ref{fig6} and \ref{fig8} and are classified to their
nearest MIR spectral type. The galaxies marked by $^{*}$ have not been
previously classified as AGN-dominant, but have MIR characteristics
which our diagnostics suggest as those of an AGN.
\end{table*}

\subsection{MIR emission of ``normal'' galaxies: early to late type}

The MIR spectra of elliptical galaxies, usually poor in cold gas and
dust, are produced primarily by their evolved stellar population, and
can be shown to resemble a blackbody continuum with temperatures
ranging from 4000 to 6000K (\cite{Madden1}, \cite{Boselli2}). This
stellar emission accounts for most of the MIR emission of some
elliptical galaxies.  However, a non-negligible fraction of elliptical
galaxies shows emission from UIBs and hot dust (\cite{Knappa},
\cite{Knappb}, \cite{Madden2}), as well as emission from non-thermal
sources.

\begin{figure}[!t]
\hspace{-8mm}
\resizebox{9.7cm}{!}{\includegraphics{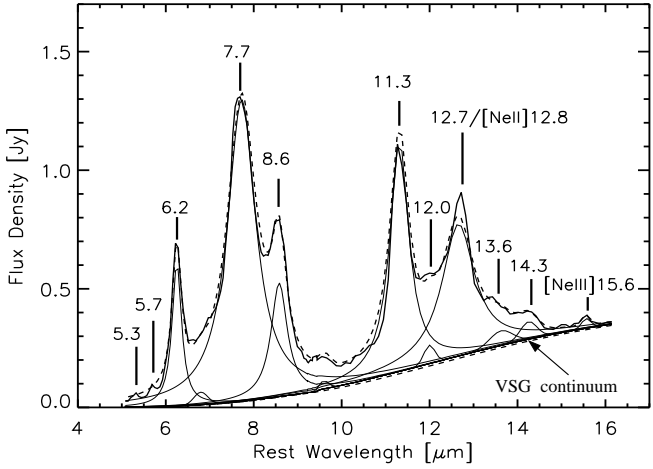}}
\caption
{Typical spectrum of a quiescent star forming region. The MIR spectrum
(5-16\,$\mu$m) of a disk region of M82 away from the central starburst
(thick solid line), chosen for its high signal to noise ratio, is
compared to a modeled spectrum (dashed line) including a continuum and
UIBs. The dust continuum due to the VSGs is modeled by a black-body curve
(T=200 K) reddened by a dust absorption law with A$_{V}$=6\,mag
(\cite{Dudley1}). The UIBs marked with the rest wavelength of their
peak emission are modeled by lorentzian profiles (thin solid lines)
(cf. \cite{Boulanger1}).  Note the faint [NeIII] emission and the
probable contamination of the [NeII]12.8\,$\mu$m line with the
12.7\,$\mu$m UIB feature. Faint UIBs are also detected at 5.3, 5.7,
12.0, 13.6, and 14.3\,$\mu$m as in many other galaxies of our sample.}
\label{fig1}
\end{figure}

The MIR spectra of spiral galaxies, rich in gas and dust, are largely
dominated by  the UIBs. The observed UIB features show little
spectral variations, and the total broad band LW3(12-18\,$\mu
m$)/LW2(5-8.5\,$\mu$m) flux ratio is close to 1 (\cite{Boselli2},
\cite{Roussel1}, \cite{Roussel2}).  This ratio has been proposed as an
indicator of the fraction of the VSG continuum to the UIB feature
emission. As an example of the ``typical'' integrated MIR emission of
normal spiral galaxies, we present in Figure~\ref{fig1} the spectrum
of a region in the disk of M82, 45\,arcsec away from the central
starburst. The usual UIB features, fitted by lorentzian profiles, are
prominent. In addition, we notice the presence of several faint
UIBs at 5.3, 5.7, 12.0, 13.6 and 14.3\,$\mu$m (\cite{Verstraete},
\cite{Beintema}, \cite{Sloan}). Similar MIR spectra are also found
in regions of the Galactic disk (\cite{Cesarskyb}, \cite{Boulanger2}).

\subsection{MIR emission of starburst galaxies.}

\begin{figure}[!t]
\hspace{-9mm}
\resizebox{9.7cm}{!}{\includegraphics{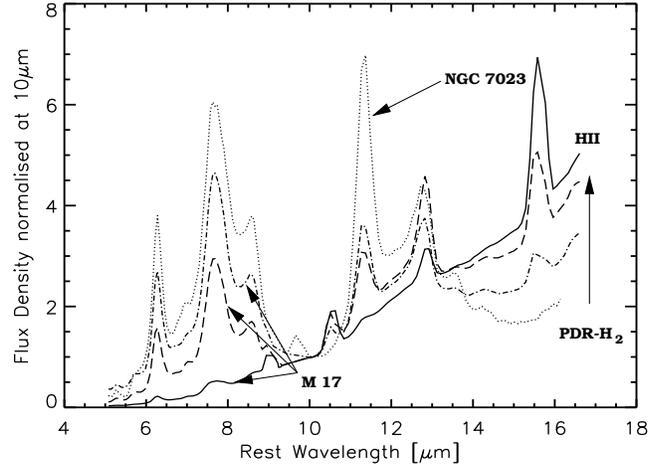}}
\caption
{MIR spectra showing the variation of the SED from a pure HII region,
depicted by a solid line (close to OB stars in M17, \cite{Cesarskyc}),
to an isolated PDR indicated by a dotted line (NGC\,7023, 
\cite{Cesarskyb}). Each spectrum was normalized to the continuum at
10\,$\mu$m. The evolution is indicated by the vertical arrow which
shows how the spectrum changes as we move from the outskirts of the
HII region (dotted line) towards its core and near the OB stars
(dashed and solid lines). Note how the strength of the UIBs
progressively diminishes and how the continuum emission increases.}
\label{fig2}
\end{figure}

The MIR spectra of galaxies with evidence of starburst activity have
distinct features compared to normal spiral galaxies. Their most
prominent characteristic is the presence of a very steeply rising
continuum at 12-16\,$\mu$m from the VSGs.  In starburst regions, this
continuum is primarily produced in HII regions, tracing regions of
massive star formation activity. The hard radiation field from young
stars also excites ionic lines from [ArII], [ArIII], [SIV], [NeII] and
[NeIII] which can be seen in the MIR spectra of starbursts.  We
display an example of this continuum in Figure \ref{fig2} (solid line)
using the spectrum near a pure HII region observed in M17, where the
MIR emission is almost completely dominated by the VSG continuum
(\cite{Cesarskyc}).  The weak intensity of UIBs is interpreted as a
consequence of the destruction of their carriers. However, since the
UIBs originate from diffuse regions as well as PDRs surrounding HII
regions, where they peak (\cite{Cesarskyb}, \cite{Verstraete},
\cite{Mattila1}, \cite{Tran}), strong UIBs can be detected in the MIR 
spectrum of embedded starburst regions.  Consequently, this depletion
of the UIB carriers will be further enhanced relative to the observed
relative increase of the MIR continuum in starburst regions (see
spectrum of Knot\,A in the Antennae Galaxies, \cite{Mirabela}).
Therefore, the LW3(12-18\,$\mu$m) to LW2(5-8.5\,$\mu$m) flux ratio
increases in strong starburst environments. One should note that the
overall MIR spectral shape can be also considerably affected by the
high extinction often found in starburst regions. In particular the
silicate band centered at 9.7\,$\mu$m can suppress the strength of
UIBs at 11.3, 8.6 and 6.2\,$\mu m$ relative to that at 7.7$\mu$m
(\cite{Lutz3}, \cite{Rigopoulou}) as seen in the spectrum of Arp\,220
(\cite{Charmandaris1}).

\subsection{MIR emission of AGNs.}

Several studies have already shown that MIR spectra in AGNs
 also present weak UIBs (\cite{Roche}, \cite{Genzel},
\cite{Schulz}, \cite{Rigopoulou}, \cite{Dudley2}).  
This effect is demonstrated in the spectrum of the nearest AGN of our
sample located in the radio galaxy Centaurus A (CenA, NGC\,5128,
\cite{Mirabelb}).  

In the MIR observations of CenA, we have sufficient spatial resolution
to disentangle the emission of the central regions near the AGN from
that due to star forming regions of the galactic disk.  The absence of
UIBs in the central 5\,arcsec ($\sim$ 100\,pc) possibly indicates the
destruction of their carriers by the intense UV-X-ray radiation field
from the AGN (\cite{Leger2}, \cite{Allain}). Moreover, there is a
noticeable continuum at short wavelengths (3-10\,$\mu$m) commonly
attributed to hot dust, associated with the torus of molecular gas
proposed in the unified model (\cite{Pier}, \cite{Granato},
\cite{Murayama}). The alternative interpretation of supernova remnants
(SNRs) as the physical mechanism explaining the AGN phenomenon
(\cite{Terlevich}) is not supported by the MIR observations. Nearby
SNRs, such as Cassiopeia A, Kepler, the Crab, RCW 103 and IC443 are
all characterized by strong ionic and/or molecular line emission and a
faint continuum (\cite{Douvion}, \cite{Oliva}).  These characteristic
features are not found in our integrated MIR spectra of galactic
nuclei which are known to harbor AGNs.  The MIR spectrum of AGNs is
flatter compared to that of pure HII regions (see the AGN and HII
spectra in Figure
\ref{fig5}). This is in agreement with IRAS observations where
infrared spectra from active galaxies are generaly significantly
flatter with the peak of emission shifted towards the MIR
(\cite{Grijp}).  Near an AGN the radiation field can heat dust up to
evaporation temperatures of $\sim$\,1000K for silicate and
$\sim$\,1500K for graphite. As a result, the dust continuum emission
becomes prominent at short wavelengths (3-6\,$\mu$m) in contrast to
emission from star forming regions which appears at longer wavelengths
(e.g. \cite{Barvainis}).  Such a continuum is found in the center of
all galaxies in our sample known to be hosting an AGN such as
NGC\,1068, NGC\,6814 or NGC 3147 (\cite{Laurent1}).  Highly ionized
species tracing the hard radiation field of the  AGN
(e.g. [NeV]14.3\,$\mu$m and [NeVI]7.6\,$\mu$m) are detected in
NGC\,1068. However, due to the low spectral resolution of ISOCAM
spectra ($\lambda / \Delta \lambda \sim$\,40), the [NeV] and [NeVI],
if they exist, are blended with the nearby UIBs in all of our galaxies 
(see also  \cite{Sturm}).

\begin{figure}[!t]
\hspace{-8mm}
\resizebox{9.8cm}{!}{\includegraphics{8899.f3}}
\caption
{Upper left panel\,: A 7\,$\mu$m map of the dust emission in Cen\,A
(from \cite{Mirabelb}).  Upper right panel\,: The central region of
Cen\,A as seen in the band 7-8.5\,$\mu$m which traces the most intense
UIB emission at 7.7\,$\mu$m. Due to a good spatial resolution at
7\,$\mu$m (FWHM\,=\,5\,arcsec, i.e. 100\,pc at 3.5\,Mpc), we can
separate the nuclear region from the disk.  Middle panel\,: The lower
curve represents the MIR spectrum originating only from the nuclear region
(100\,pc in diameter) whereas the upper curve shows the spectrum
integrated over a larger region (800\,pc in diameter) which includes both
disk structures and the nucleus. The silicate absorption  cannot
be well estimated from the global spectrum but is clearly detected in
the AGN continuum. Lower panel\,: The AGN spectrum (left) represents
45\,$\%$ of the energy between 5 and 16\,$\mu$m and it contributes
more than 90\,$\%$ at wavelengths between 3 and 6\,$\mu$m. No emission
associated with star forming regions is detected between 3-4\,$\mu$m
(lower right panel, ISOCAM SW channel).  While UIBs dominate the MIR
emission of the disk, they are almost absent in a region of 50 pc
radius around the AGN.  The higher [NeIII]/[NeII] ratio detected in the
AGN (lower left panel) further indicates the presence of a hard radiation
field.}
\label{fig3}
\end{figure}

\section{MIR diagnostics to distinguish AGNs, starburst and quiescent star forming regions.}

Our diagnostics are based on the assumption that the integrated MIR
emission in galaxies can be represented by a sum of contributions
originating from (1) regions where the dust is predominantly heated by an
AGN and (2) regions where star formation is the main source of energy.
As mentionned in sections 3.1 and 3.2, star forming regions can also
be divided in two classes: relatively quiescent regions where most of
the MIR emission originates from the PDRs, and starburst-like regions
where continuum emission at 15\,$\mu$m produced in the HII regions is
dominant. We therefore propose to categorize the spectra from our
sample using empirical criteria chosen to separate the distinct UIB
and MIR continuum behaviours. To further examine the relative
proportion of each contribution, we also compare our spectra with
three templates, namely, AGNs, pure HII regions and isolated
PDRs. The relative proportions of these templates required to
reproduce a given MIR spectrum allow us to classify the observed
region as MIR spectral type dominated by AGN, starburst or relatively
quiescent regions.

\subsection{MIR templates}

To construct the templates of each component, we selected three ISOCAM
CVF spectra dominated by each of these physical processes. These
template spectra are presented in Figure \ref{fig5}a.

The strong continuum at 14-15\,$\mu$m observed in starburst galaxies
was modeled by a typical pure HII Galactic region spectrum from M17
(\cite{Cesarskyc}). UIBs are absent or faint (upper panel of Figure
\ref{fig5}a) and the [ArIII](8.9\,$\mu$m), [SIV](10.5\,$\mu$m),
[NeII](12.8\,$\mu$m) and [NeIII](15.6\,$\mu$m) are present. A typical
starburst spectrum can be usually decomposed using the VSG continuum
with some UIB emission (\cite{Tran}).  Since UIBs originate mainly
from PDRs or diffuse regions, we used as their template, an isolated
PDR spectrum from the galactic reflection nebula NGC\,7023
(\cite{Cesarskyb}) (see the lower right part of the Figure
\ref{fig5}a).

The AGN component  was  characterised by the MIR spectrum of the central
region of CenA (see Fig. \ref{fig3} and the lower left part of the
Figure \ref{fig5}a), in which UIBs are absent as in the HII region
template. However, in contrast to the HII region template spectrum,
the continuum at short wavelengths (5-10\,$\mu$m) is prominent in the
AGN spectrum. In addition, the equivalent width of emission lines of
[NeII] and [NeIII] is small, perhaps due to the overwhelming continuum
contribution. Note that extinction affects the AGN template and is
greater in Cen\,A than in NGC\,1068 (see Fig. \ref{fig10}).

\begin{figure}[!t]
\hspace{-7mm}
\resizebox{9.5cm}{!}{\includegraphics{8899.f4}}
\caption
{The relative strength of the UIB(6.2\,$\mu$m) feature is estimated by
dividing the spectrum between 5.1 and 6.8\,$\mu$m in two different
parts. One is the integrated flux above the line between 5.9 and
6.8\,$\mu$m which is attributed to the UIB at 6.2\,$\mu$m and traces
essentially the PDRs in the star forming regions. The second is the
integrated flux under the line which is mainly attributed to the AGN
continuum and/or wing of the UIB at 7.7\,$\mu$m. The feature to
continuum indicator is used to estimate the relative contributions to
the heating of the dust from quiescent star forming regions compared
to the one from an AGN. We measure the ``hot continuum''
(5.1-6.8\,$\mu$m) which may contain an AGN contribution and compare it
to the ``warm continuum'' (14-15\,$\mu$m) produced by both star
forming regions and AGNs. As a consequence, the warm to hot continuum
ratio can be used to quantify the relative contribution of a starburst
continuum and an AGN in the MIR.}
\label{fig4}
\end{figure}

\subsection{MIR indicators: [HII]/[AGN] and [PDR]/[AGN,HII]}

\begin{figure*}[!t]
\resizebox{18cm}{!}{\includegraphics{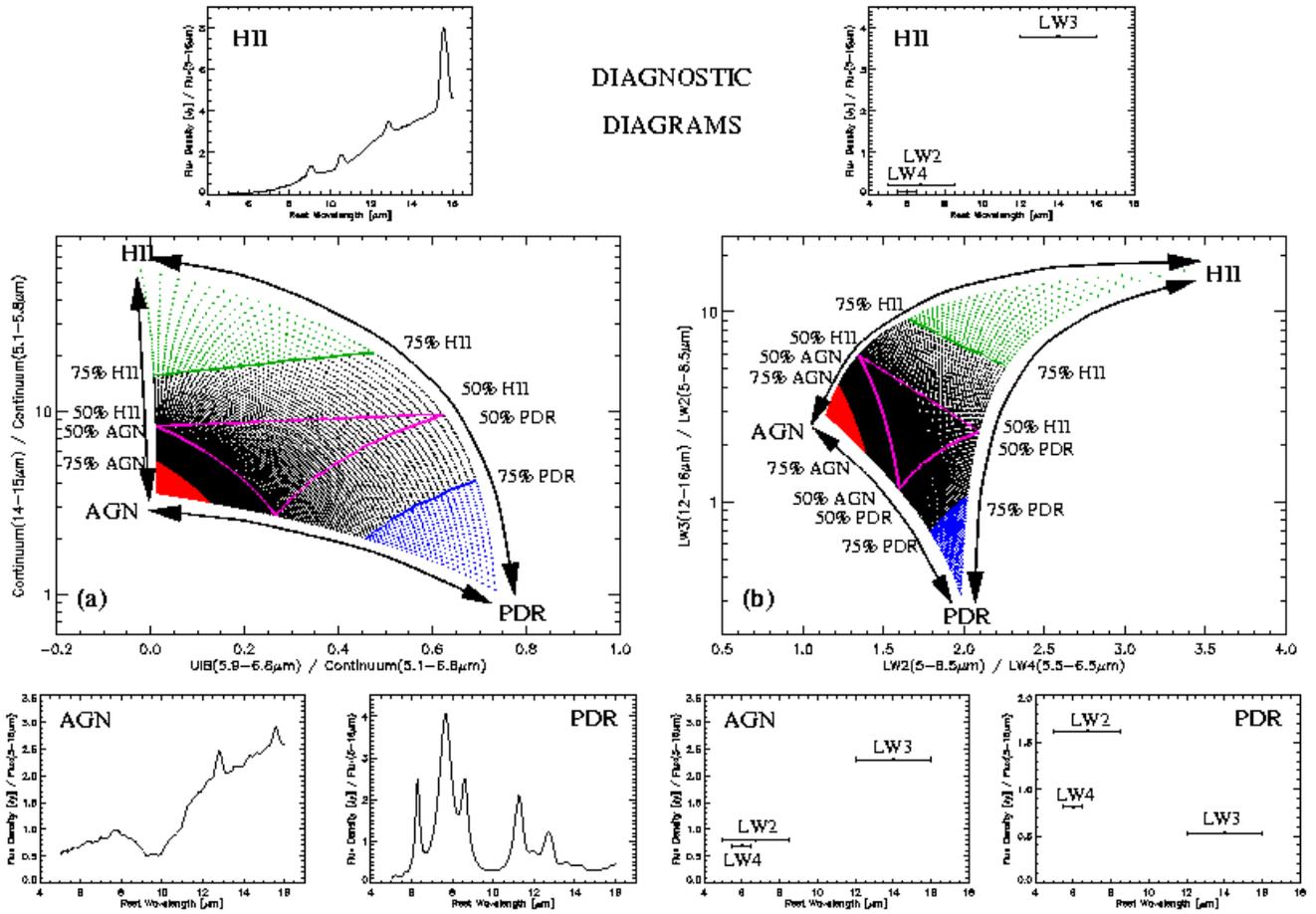}}
\caption
{a) Diagnostic diagram based on CVF spectra. b) Diagnostic diagram
based on broad band filters.  Three distinct areas can be defined. In
the upper corner, we select spectra dominated by massive HII regions
such as those found in starbursts. On the lower left, AGN spectra are
dominant in a very small region, and finally, PDR spectra fall in the
lower right part. The solid lines and the associated percentages
indicate a constant fraction of one component along each line (see
section 4.3).}
\label{fig5}
\end{figure*}

As each MIR template has distinct characteristics, we can quantify
their relative contribution to the integrated spectrum. We have used
two indicators based on the UIB strength and the MIR continuum. As an
indicator of the relative HII region and AGN contribution, we use the
ratio of ``warm'' (14-15\,$\mu$m) to ``hot'' (5.1-6.8\,$\mu$m)
continuum (Fig. \ref{fig4}). As this ratio decreases, the presence of
an AGN becomes more evident due to the increase of the relative
importance of the ``hot'' continuum.  The UIB(6.2\,$\mu$m) to
continuum (5.1-6.8\,$\mu$m) flux ratio can be used to identify the
relative contributions of the quiescent star forming regions to the
starburst and/or AGN MIR emission. This indicator is very similar to
the one proposed by Genzel which is based on the strength of the
7.7\,$\mu$m feature (\cite{Genzel}, \cite{Lutz3}).  Our estimate of
the UIB flux has the advantage to be less affected by the strong
silicate extinction at 9-11\,$\mu$m while the continuum at 5-7\,$\mu$m
is also less contaminated by the VSG continuum. The effect of a strong
absorption on the AGN continuum could artificially enhance the
presence of a UIB at 7.7\,$\mu$m but not at 6.2\,$\mu$m. Nevertheless,
the choice of the 6.2\,$\mu$m which is weaker than the 7.7\,$\mu$m
feature necessitates spectra with good signal to noise ratio. This is
the case in our sample of spectra where the UIB at 6.2\,$\mu$m is very
well detected (Figures \ref{fig3} and \ref{fig7} or the spectra
in the Antennae in \cite{Mirabela}, for some examples). Since our
primary goal is to establish the absence of UIBs as well as the
presence of a well detected AGN continuum, at short MIR wavelengths, a
criterion based on the UIB(5.9-6.8\,$\mu$m)/Continuum(5.1-6.8\,$\mu$m)
ratio applied on high signal to noise spectra gives a more accurate
diagnostic for revealing the presence of an AGN. As we show in the
following sections, the relative variation of our UIB/Continuum at
6\,$\mu$m is reliable in separating a featureless continuum from UIB
spectra even if our 6\,$\mu$m continuum measurements may still be
slightly affected by emission coming from the wing of the 7.7\,$\mu$m
UIB (see Fig. \ref{fig4}).

\subsection{Diagnostic diagram}

Having defined the three MIR templates, we construct a complete
library of composite spectra where the contribution of each template
can vary between 0 and 100\,$\%$.  On these composite spectra, we can
measure the diagnostic ratios presented above and plot them on a
2-dimensional diagram (see Fig. \ref{fig5}). Three main regions can be
identified in the corners of this diagram where the signatures of
AGNs, starbursts, or quiescent star forming regions dominate the total
spectrum.

\begin{figure*}[!t]
\hspace{-10mm}
\resizebox{19.2cm}{!}{\includegraphics{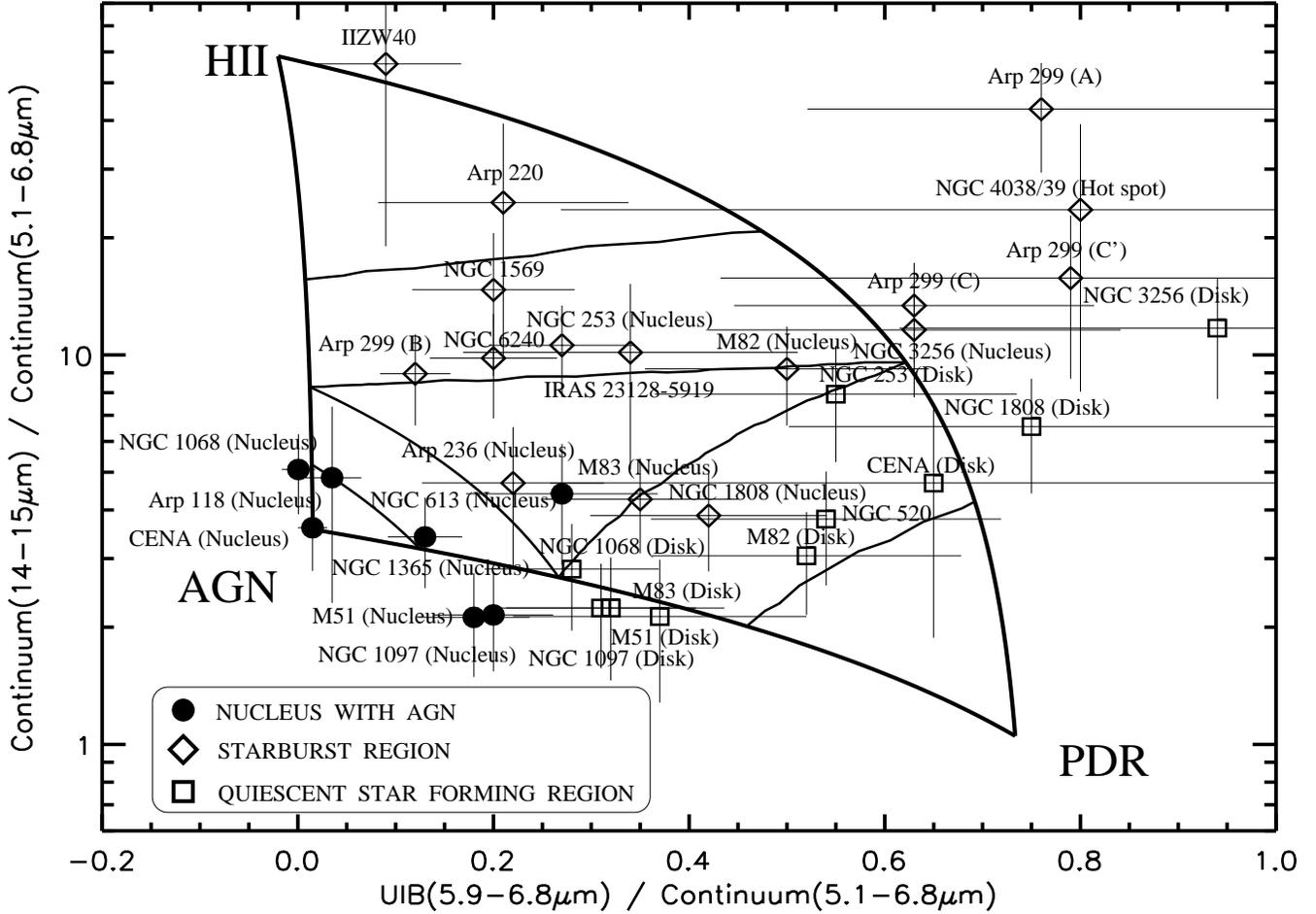}}
\vspace{-7mm}
\caption
{Diagram based on CVF diagnostics for 33 spectra using the notation of
Table 1.  The error bars are estimated from the rms maps at 1\,$\sigma$
for each wavelength along the CVF spectra, added to a systematic error
of 10$\%$ due to the transient correction method. Note that several
spiral and blue compact dwarf galaxies included in this figure are not
listed in the Table \ref{sample} since they belong to other samples
(\cite{Madden2}, \cite{Roussel1}). The circles mainly located in the
bottom left part of the diagram represent the central regions known to
contain an AGN.  The galaxies hosting starburst activity are in the
upper part and are marked with diamonds. The spectra of quiescent star
forming regions are found to lie close to the PDR region and are denoted
with squares. The curves demarcate the AGN, PDR and HII spectral type
according to Figure \ref{fig5}a.}
\label{fig6}
\vspace{0mm}
\end{figure*}

The above diagram can be applied only in the galaxies where
MIR spectra are available. However, one  may expand it to
cases where only broad band MIR images are available. This can be
done by using our library of spectra to estimate their corresponding
fluxes through the ISOCAM filters that isolate as closely as possible
the regions identified in Figure~\ref{fig4}. The filters used are the
LW2(5-8.5\,$\mu$m), LW3(12-18\,$\mu$m) and LW4(5.5-6.5\,$\mu$m).
As we will show in the following sections the broad band diagnostic
gives results comparable to those obtained
with the CVFs. The equivalent fluxes of our CVF templates through the three
broad band filters are presented in Figure \ref{fig5}b. Using the same
reasoning as in Figure \ref{fig5}a, we estimate the relative strength
of UIBs with the LW2(5-8.5\,$\mu$m)/LW4(5.5-6.5\,$\mu$m) flux ratio
while the ratio of the LW3(12-18\,$\mu$m)/LW2(5-8.5\,$\mu$m) emission
provides an estimate of the VSG contribution relative to the UIB
and/or hot dust emission at short wavelengths. Those two diagnostic
diagrams have in common the capacity to separate individualy regions
dominated by AGN, starburst or quiescent star forming regions. The
robustness of the two criteria against observed MIR spectra of
galaxies is examined in the following section.

\begin{figure*}[!t]
\vspace{-0cm}
\hspace{-3mm}
\resizebox{18cm}{!}{\includegraphics{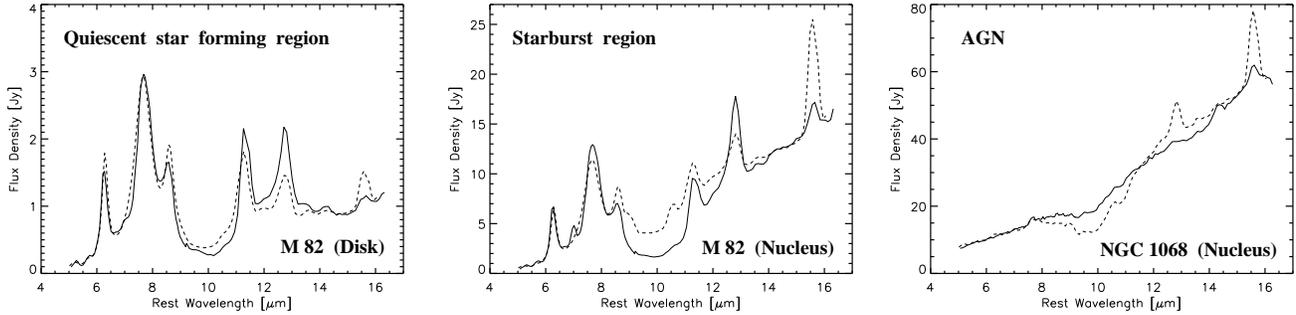}}
\vspace{-2mm}
\caption{Examples of spectra (solid lines) classified as dominated by star 
forming regions (the disk and the nucleus of M82) and by an AGN (the
nucleus of NGC\,1068) are shown with their corresponding model based
on a combination of templates (dashed line). For each spectrum, we
find the appropriate combination of templates (8$\%$AGN, 50$\%$HII,
42$\%$PDR for the nucleus of M82; 8$\%$AGN, 15$\%$HII, 77$\%$PDR for
the disk of M82 and 77$\%$AGN, 23$\%$HII, 0$\%$PDR for the nucleus of
NGC\,1068) using the location of those sources in the CVF diagram 
(see Fig. 6). 
Each composite spectrum has been normalised at 14-15\,$\mu$m.}
\label{fig7}
\end{figure*}

\begin{figure*}[!t]
\hspace{-11mm}
\resizebox{19.2cm}{!}{\includegraphics{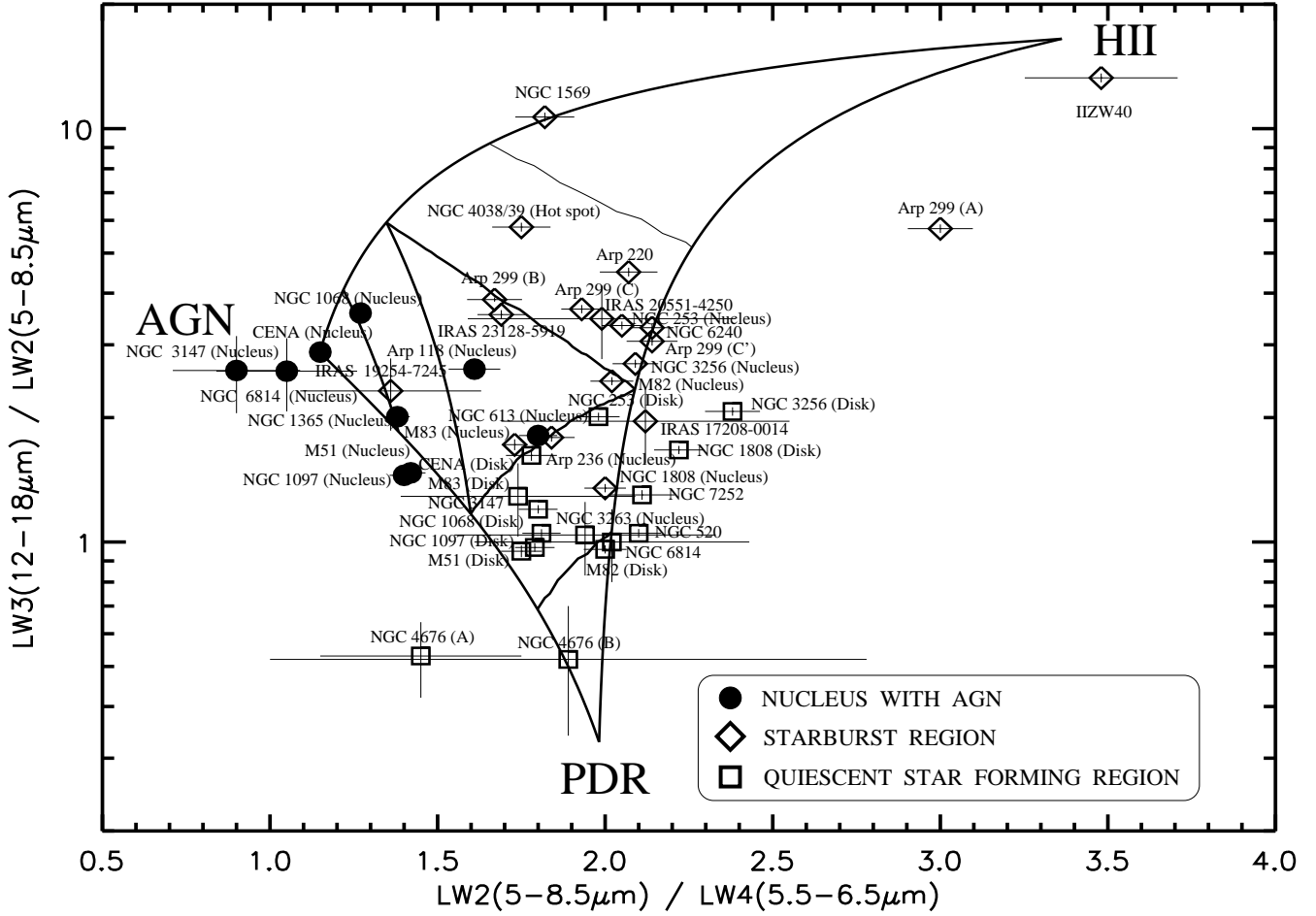}}
\vspace{-5mm}
\caption
{Diagram based on broad band diagnostics for 45 spectra using the
notation of Table 1.  The error bars are estimated from the rms map at
1\,$\sigma$ for CVF spectra or broad band observations added to a
systematic error of 10$\%$ due to the transient correction
method. Note that
several spiral and blue compact dwarf galaxies not listed in the Table
\ref{sample} are from other samples (\cite{Roussel1},
\cite{Madden2}). The galactic central regions (size of several kpc)
hosting an AGN are all located in the left part of the diagram (marked
with circles).  The galaxies dominated by a starburst activity are in
the upper part due to their higher LW3/LW2 ratio (marked with
diamonds).  The spectra of quiescent star forming regions lie close to
the PDR area (marked with squares).  The curves demarcate the
AGN, PDR and HII spectral type according to Figure \ref{fig5}b.}
\label{fig8}
\end{figure*}

\subsection{Application of the diagnostic}
Our MIR indicators were applied to the galaxies of our sample (see
Table \ref{sample}), in which we also included a few irregular
galaxies such as the blue compact dwarf galaxies IIZw40 and NGC\,1569
(\cite{Madden2}) and barred spiral galaxies NGC 1097 and NGC 1365
(\cite{Roussel1}) in order to cover the diagram completely. In cases
where we had adequate spatial resolution, we extracted different
physical regions in the same galaxy. The results are presented in
Figure \ref{fig6}. Galaxies commonly classified as starbursts, such as
Arp\,220 and Arp\,299(A), fall toward the top part of the diagram
implying the presence of a large fraction of starburst regions
compared to quiescent star forming regions.  A few nearby galaxies
with known AGNs are located in the bottom left part of the diagram
along with our AGN template, the nucleus of Cen\,A. Regions of
moderate star formation observed in galactic disks are situated in the
region of the diagram where we expect to find a dominant PDR
contribution.  In Figure \ref{fig7}, we present three typical spectra
from our sample dominated by UIBs (the disk of M82), an HII continuum
(the nucleus of M82) as well as an AGN (the nucleus of
NGC\,1068). Based on their position on our diagnostic diagram in
Figure \ref{fig6}, we estimate the corresponding fraction of MIR
templates according to the Figure \ref{fig5}a.  This is a first order
approximation since we do not take explicitly into account the effects
of the 9.7 \,$\mu$m silicate absorption band on our template
spectra. Consequently, the observed difference between model and
spectra in the range of 9 to 11\,$\mu$m is probably due to the strong
silicate extinction in embedded starbursts or AGNs.  We do not attempt
to model the emission from the [NeII](12.8\,$\mu$m) and
[NeIII](15.6\,$\mu$m) lines which are independant of the dust emission
properties. Nevertheless, the overall agreement presented in Figure
\ref{fig7} is sufficient for modeling the general shape of spectra.

To extend our diagnostics in the case where spectra are not available,
we present in Figure \ref{fig8} the diagnostic diagram using broad
band filters\footnote{For galaxies of our sample for which we had only
CVF spectra, we calculated the equivalent broad band filter fluxes
taking into account their filter transmission. Note that since the CVF
ends at 16\,$\mu$m to match its equivalent LW3 filter to the ISOCAM
standard LW3(12-18\,$\mu$m) filter, we have normalized the
transmission of the latter between 12 and 16\,$\mu$m.}.  As expected,
the AGN candidates appear in the left part of this diagram.  For
LW3/LW2 ratios close to 1, PDRs dominate the MIR emission in our
sample as it was the case for normal spiral galaxies (\cite{Boselli1},
\cite{Roussel1}). For LW3/LW2 ratios between 1 and 6, both an AGN and
a star forming region signature may appear in our spectra (e.g. the
nucleus of NGC\,253 and NGC\,1068 in Fig.~\ref{fig8}).

We note, however, that the LW2/LW4 flux ratio is well adapted to
estimate the presence of an AGN contribution. For LW2/LW4 ratios lower
than 1.5, we clearly see that the MIR emission is dominated by the AGN
whose strong continuum and nearly absent UIBs, contribute to decrease
this ratio. For larger LW2/LW4 ratios, emission is dominated either by
quiescent star forming regions or by starburst regions (see
Fig. \ref{fig8}). Applying this criterion to the nucleus of M82, we
clearly see that it is classified as a starburst while the nucleus of
NGC\,1068 has a typical AGN signature. Athough this diagram uses only
three broad band filters, the MIR classification is in good agreement
with results given by the diagnostic diagram based on full spectra
(see Table \ref{sample}).

Moreover, the diagnostic diagrams of Figures \ref{fig6} and
\ref{fig8} suggest that the well-known ultra-luminous galaxies
such as IRAS 23128-5919 and NGC\,6240 have MIR spectra which are
overwhelmed by a strong starburst signature. In the ``Super Antennae''
(IRAS 19254-7245), which harbors a Seyfert 2 nucleus (\cite{Mirabel2}),
more than $\sim$70$\%$ of its MIR flux originates from the AGN
(\cite{Laurent6}).

\begin{figure*}[!t]
\hspace{-5mm}
\resizebox{18.5cm}{!}{\includegraphics{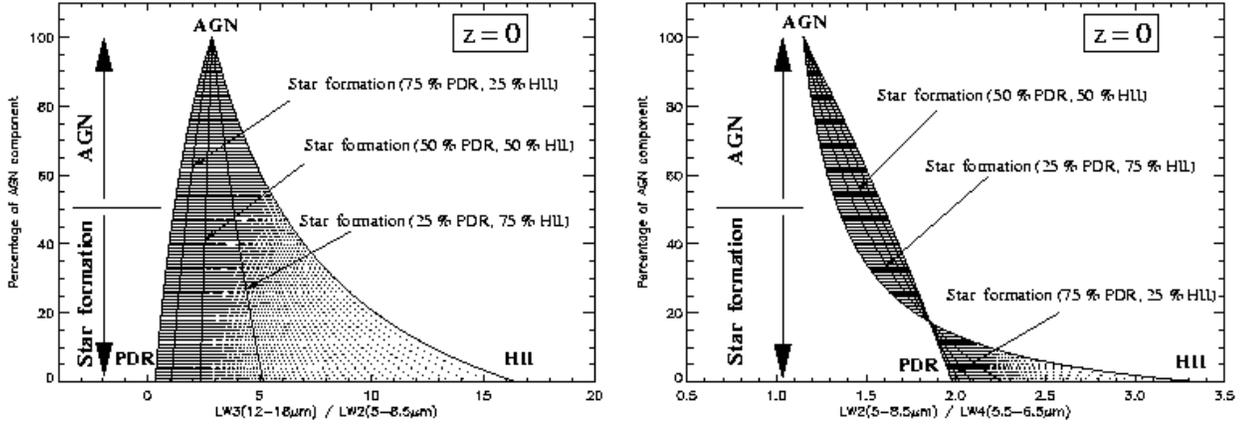}}
\caption
{ a)In the above figure we display the variation of the AGN-Starburst
fraction (indicated in the y-axis) as a function of
LW3(12-18\,$\mu$m)/LW2(5-8.5\,$\mu$m) flux ratio based on the three
templates of our diagnostic for galaxies of the local universe.  We
divide our library of composite spectra (see Fig. \ref{fig5}) into one
component attributed to star formation activity (including both the
PDR/diffuse and HII region templates) and a second one to AGN.  For a
given LW3/LW2 ratio and a selected type of star fomation mixture (as
indictated by the solid curves of different PDR and HII fractions) one
can calculate the percentage of the AGN component in the spectrum.  b)
Same as in a) but for the LW2(5-8.5\,$\mu$m)/LW4(5.5-6.5\,$\mu$m) flux
ratio. Note how the use of just the LW3/LW2 ratio is insufficient for
detecting AGNs. This is not the case for the LW2/LW4 ratio which is
relevant to select AGNs from starbursts (see section 4.4).}
\label{fig9}
\end{figure*}

One can also use this technique to discover up to now unclassified AGN
hidden by a large column density of dust. Such a candidate is Arp\,236
(IC\,1623, VV\,114), an infrared luminous system (L$_{IR}$ =
10$^{11.62}$ L$_\odot$), classified as an early-stage gas-rich merger
(M(H$_2$) = 5.1$\times$10$^{10}$ M$_\odot$, \cite{Yun}), and composed
of several nuclei and/or compact starburst regions.  Previous
near-infrared and radio studies showed no evidence for an AGN
(\cite{Knop}, \cite{Doyon}). However, the MIR signatures of the
eastern region (knot A, \cite{Doyon}) place it near the AGN locus in
our CVF diagram (Fig. \ref{fig6}) and, therefore, may contain a
non-negligible fraction of AGN contribution in its MIR
spectrum. Nevertheless, one should note that the possible AGN
contribution in Arp\,236 is not revealed with the LW diagnostic
(Fig. \ref{fig8}) which is less accurate compared to the CVF
diagnostic.  Knot A is the brightest source in the near-infrared. It
is also redder and more compact than all other sources in the system
(not resolved at 3.7\,$\mu$m with a FWHM of 0.3\,arcsec,
\cite{Knop}). We estimate that $\sim$\,40\,$\%$ of the MIR emission in
the region can originate from an optically obscured AGN. A MIR
spectral classification based on our method is presented in the last
column of Table \ref{sample} for all of our sample.

\section{Discussion}

Our diagnostic diagrams provide a new tool allowing us to identify an
AGN or star formation (starburst and/or quiescent star forming 
regions) signature in the integrated MIR spectrum of a galaxy.

Although self consistent, our method is based on data from a rather
diverse sample. To further test its validity, we compared it with the
MIR diagnostics developed by Genzel et al. (1998a) on a complete
sample of ULIRGs, using ISOPHOT-S. Using the published MIR spectra of
the brightest sources in the Genzel sample, we calculated the strength
of the UIB at 6.2\,$\mu$m in these spectra, as well as the equivalent
LW2(5-8.5\,$\mu$m)/LW4(5.5-6.5\,$\mu$m) ratio since this is the
principal discriminator between an AGN and star forming regions (see
Fig. \ref{fig9}). The classification of galaxies obtained based on our
method is in complete agreement with that of Genzel et
al. (1998a). IRAS 23060+0505, IRAS 19254-7245, Mrk 231 and Mrk 273,
with an LW2/LW4 ratio of 1.16, 1.36, 1.42 and 1.60 respectively, are
classified by both methods as galaxies containing a significant AGN
contribution in the MIR.  Furthermore, in the wavelength range of
ISOCAM (5-16\,$\mu$m), we can better constrain the nature of the
continuum at 7\,$\mu$m between dust and stellar contribution using the
14-15\,$\mu$m flux, since the LW3(12-18\,$\mu$m)/LW2(5-8.5\,$\mu$m)
flux ratio is greater than 3 for the dust continuum found in AGNs or
HII regions and less than 0.4 for the stellar continuum (Boselli et
al. (1997).  We are also able to distinguish whether a featureless
continuum is due to an AGN or a pure HII region (e.g. IIZw40).

Selective absorption by amorphous silicates, centered at 9.7 and 18
$\mu$m, can play a crucial role in obscuring emission over much of the
MIR wavelengths. AGNs embedded in a large amount of dust could still
remain undetected by our diagnostic. Studying a sample of 28 Seyfert 1
 and 29 Seyfert 2 with ISOPHOT-S, Schulz et
al. (1998) have already shown that the high absorption in Seyfert 2
galaxies blocks a large fraction (90$\%$ on the average) of the MIR
continuum from the AGN inner torus. They do detect the
silicate feature in emission at 9.7\,$\mu$m which is in favour of a
moderately thick torus model (A$_{V}$$\sim$100, \cite{Granato}) and
rules out models with very large optical depths (A$_{V}$$\sim$1000,
\cite{Pier}). In our sample, among all nuclei (diameter\,$<$\,9$''$) which are 
classified as AGNs based on our MIR diagnostics (see Fig. \ref{fig8}),
five nuclei are optically selected as Seyfert 2.  Provided sufficient
spatial resolution is available, we can detect the hot dust continuum
associated with the AGN and we are able to disentangle its
contribution to the MIR emission of the galaxy.  If those nuclei were
observed with the ISOPHOT-S aperture of 24$\times$24\,arcsec$^2$, we
would detect the presence of an AGN only in the two nearest Seyfert 2
of our sample NGC\,1068, CenA and in NGC\,6814 which is a Seyfert
1. The dilution effect combined with the optical depth of the AGN
torus would lead to an underestimation of the number of detected AGNs.
The MIR continuum associated with AGNs is not completely suppressed
behind the dusty torus and may be observed using sufficient spatial
resolution ($\sim$5\,arcsec) observations on nearby galaxies
(D$<$\,50\,Mpc with H$_{0}$=75\,km\,s$^{-1}$Mpc$^{-1}$, see
\cite{Laurent3}). The detection of the AGN continuum in Seyfert 2 is
also in better agreement with a dusty torus model producing moderate
absorption (A$_{V}$$\sim$100) as in the model of Granato et
al. (1997).

\begin{figure}[!t]
\hspace{-7mm}
\resizebox{9.5cm}{!}{\includegraphics{8899.f10}}
\caption
{The effects of absorption on AGN spectra characterised by a
strong continuum at short wavelength associated with an absence of
UIBs. In small panels, we display the nuclear spectra of NGC
1068 and CenA as well as a best fit based on a simple screen model
(with the dust absorption law of \cite{Dudley1}) applied to a power law of
spectral index $\alpha$. The corresponding visible absorptions for the fits
are 7 and 20 mag respectively. The large diagram presents
how the same power law spectrum ($\alpha$=1.7) affected when absorption 
ranges from A$_{V}$=0 to 100.}
\label{fig10}
\end{figure}

In Figure \ref{fig10}, we note that the increase of the absorption in
an AGN spectrum leads to a decrease of the hot continuum used in our
diagnostic.  This further suggests that the ``true'' AGN fraction
would be underestimated and our detections must be considered as a
lower limit.  An intrinsically low AGN contribution to the MIR
spectrum of a source would be difficult to distinguish from the star
forming emission at 6-7\,$\mu$m composed by UIBs and the VSG
continuum. Consequently, we consider that an AGN is detected by our
diagnostic when the estimated fraction of the AGN emission is larger
than the contribution of starburst or quiescent star forming regions.

\begin{figure*}[!t]
\hspace{-5mm}
\resizebox{18.5cm}{!}{\includegraphics{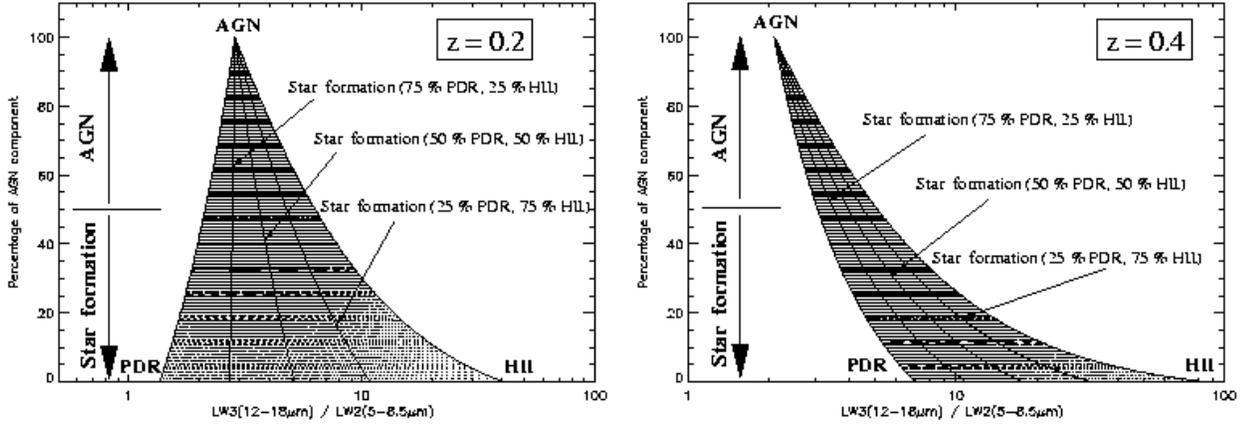}}
\caption{Variation of the AGN-Starburst fraction at different 
redshifts (z=0.2 on the left and z=0.4 on the right) as a function of
LW3(12-18\,$\mu$m)/LW2(5-8.5\,$\mu$m). We use the same notation as in
Figure \ref{fig9} for the contribution of the star formation
component.  The percentages indicate the fraction of each contribution
in the rest-frame wavelengths.  The LW3/LW2 ratio which is degenerate
for distinguishing AGNs from starbursts at low redshifts can be used
for detecting the AGN continuum at higher redshifts (i.e. z$\sim$0.4)
 due to the K-correction effect on MIR spectra (see section 5).}
\label{fig11}
\end{figure*}

Recently several studies on the origin of the MIR emission in galaxy
clusters and deep field surveys were performed with ISOCAM using the
LW3 and LW2 broad band filters (e.g. \cite{Aussel}, \cite{Desert2},
\cite{Flores}).  Our diagnostic implies that one cannot discriminate
between an AGN and a starburst signature using just those two filters
in the local universe (see Fig.~\ref{fig9}). At high redshifts, the
two filter bandpasses sample shorter rest frame wavelength emission
and are less sensitive to the VSG continuum.  In Figure \ref{fig11},
we applied a redshift correction on our MIR templates at z=0.2 and
z=0.4, and we obtained the equivalent diagnostic diagrams derived from
observations of nearby galaxies. To extrapolate the part of the
spectrum under 5\,$\mu$m, we use a power law for the AGN template and
a black-body continuum for the pure HII region template. We also
assume that the PDR component under 5\,$\mu$m is negligible
(\cite{Lu}). At z=0.4, objects with LW3/LW2$<$3 are classified as
AGN-dominant in the MIR.  For redshifts between 0.5 and 1, the LW3
filter samples mainly UIBs (LW2 at z=0), while LW2 probes the hot dust
emission between 3 and 5\,$\mu m$ (assuming that the stellar emission
is negligible in spiral galaxies and only begins to be substantial
below 2\,$\mu$m). Consequently, low values of the LW3/LW2 flux ratios
in these surveys would suggest the presence of  an AGN.

In our diagnostic method, we have implicitly considered  that the MIR
emission from the evolved stellar population is negligible in spiral
galaxies.  Even though this may be true for galactic disks
(\cite{Roussel1}), extra caution is needed for bulges in the center of
which one finds the AGN. In that case, it is the rising slope of the
MIR spectrum which would reveal the presence of an AGN, even though we
cannot rule out completely a possible faint contamination from evolved
starlight at short wavelengths. Moreover, the fact that the core
radius of the bulge is several orders of magnitude larger than the
region responsible for the ``hot continuum'' would in principle
facilitate their separation in nearby galaxies.  For distant galaxies
where the integrated MIR spectra include the whole galactic bulge, it
is very difficult to estimate the form of the continuum below
6\,$\mu$m.  M83, a well known starburst galaxy, is such an example,
since we detect considerable continuum emission at 5-6\,$\mu$m (see
Fig. \ref{fig8}). This emission is clearly extended outside the
unresolved nuclear region and can be interpreted as stellar emission
from the stellar bar (cf. \cite{Elmegreen} and Sauvage
priv. comm.)\footnote{To further develop our diagnostic for general
cases including evolved stellar population, the K band flux at
2.2\,$\mu$m could be used to estimate the contribution of the stellar
emission observed at short MIR wavelengths. A LW2/K or LW3/K flux
ratio less than 1 would indicate a more important contribution from the
stellar component.}. Furthermore, for distant galaxies the
contribution of the star forming regions surrounding an AGN would
progressively enter in the beam and dilute the AGN MIR signatures. A
more detailed discussion on this issue is presented in
\cite{Laurent3}.

Our diagnostic method can be further expanded with the improved
performance of upcoming telescopes, and can be used as a guide in
scheduling future research programs. In addition to an increase in
sensitivity, SIRTF will provide better wavelength coverage (5-40
$\mu$m) than ISOCAM, and better spatial resolution than the ISO-SWS
observations.  In particular, SIRTF would provide a better measurement
of the continuum for distant faint sources and, calculating the depth
of both silicate absorption bands at 9.7 and 18 $\mu$m, a more precise
estimate for the absorption. An increase in spatial resolution would
also permit a significant improvement in the AGN/Starburst
diagnostics. Both the infrared instruments installed on ground based
10 meter telescopes and the Next Generation Space Telescope which will
cover the MIR spectrum, will probe regions of $\sim$10\,pcs near the
AGNs, decreasing significantly the effects of beam dilution.

\section{Conclusions}

Using our ISOCAM MIR observations we have obtained a new AGN/Starburst
diagnostic based on the strength of the UIBs at 6.2\,$\mu$m and the
MIR continuum.  We conclude that:

1) In AGN spectra (even with a faint starburst contamination), a
strong MIR continuum is present at short wavelengths between
3-10\,$\mu$m. This continuum may be attributed to very hot dust grains
directly heated by the central engine. Furthermore, the absence of
UIBs suggests that their carriers can be destroyed by the strong UV-X
ray radiation field.

2) Our MIR diagnostic diagrams can be used to unravel AGNs that are
completely obscured at optical wavelengths. For example, the nucleus
of Arp\,236 which was classified as a starburst galaxy triggered by an
interaction is likely to contain an AGN that contributes $\sim$ 40$\%$
to the MIR flux.

3) The emission from ultraluminous infrared galaxies can be dominated
either by a starburst (e.g. Arp 220) or an AGN (e.g. The Super
Antennae, IRAS 19254-7245). Nevertheless, an AGN can remain partially
hidden by the torus absorption even in the MIR which may lead to an
underestimate of its contribution to the whole spectral energy
distribution.

4) In the local universe, the LW3(12-18\,$\mu$m)/LW2(5-8.5\,$\mu$m)
ratio alone cannot be used to distinguish AGNs from
starbursts. However, for galaxies at z=0.4-1, this ratio is effective
in discriminating the AGN from the starburst contribution.

5) Applications of adapted versions of our diagnostic in future
instruments such as SIRTF and NGST, which will sample the MIR band with
higher spatial resolution and sensitivity than ISOCAM, should provide
better insights into the heating sources of the interstellar medium in
galaxies.

\begin{acknowledgements}
We thank H. Roussel, D. Tran and T. Douvion for providing data in
advance of publication as well as P.-A. Duc and D.B. Sanders for their
help on various aspects of this work. We are grateful to C. Dudley for
sending us his extinction curve.  We thank the referee, C. Lonsdale,
for useful suggestions which led to considerable improvement of this
paper. VC would like to acknowledge the financial support from a Marie
Curie fellowship (TMR grant ERBFMBICT960967).
\end{acknowledgements}

\end{document}